\documentclass[fleqn,usenatbib]{mnras}

\usepackage{newtxtext,newtxmath}

\usepackage[T1]{fontenc}

\DeclareRobustCommand{\VAN}[3]{#2}
\let\VANthebibliography\thebibliography
\def\thebibliography{\DeclareRobustCommand{\VAN}[3]{##3}\VANthebibliography}

\usepackage{amsmath}	
\usepackage{booktabs}
\usepackage{graphicx}
\usepackage[shortlabels]{enumitem}
\usepackage[countmax]{subfloat}
\usepackage{anyfontsize}
\usepackage{xcolor}
\newcommand{\rev}[1]{\textcolor{magenta}{#1}}

\usepackage{booktabs}
\usepackage{amsmath}
\usepackage{array}
\usepackage{wasysym}

\newcommand{\Htwo}{H$_{2}$}

\newcommand{\HII}{\ion{H}{ii}}

\newcommand{\Fe}{[\ion{Fe}{ii}]}

\newcommand{\SII}{[\ion{S}{ii}]}
\newcommand{\PII}{[\ion{P}{ii}]}
\newcommand{\SIII}{[\ion{S}{iii}]}
\newcommand{\OI}{[\ion{O}{i}]}

\newcommand{\OIII}{[\ion{O}{iii}]}

\newcommand{\NII}{[\ion{N}{ii}]}

\newcommand{\CI}{\ion{C}{i}}

\newcommand{\Ha}{H$\alpha${}}
\newcommand{\Hb}{H$\beta${}}

\newcommand{\pab}{Pa$\beta${}}

\newcommand{\pag}{Pa$\gamma${}}
\newcommand{\brg}{Br$\gamma${}}
\newcommand{\wha}{W$_{H\alpha}${}}
\newcommand{\wpb}{W$_{Pa\beta}${}}

\newcommand{\kms}{km~s\pwr{-1}}

\newcommand{\pwr}[1]{$^{#1}$}

\makeatletter
\newcommand*{\rom}[1]{\expandafter\@slowromancap\romannumeral #1@}
\makeatother
\makeatletter
\def\blfootnote{\xdef\@thefnmark{}\@footnotetext}
\makeatother

\newcolumntype{R}[1]{>{\RaggedLeft\arraybackslash}p{#1}}
\newcolumntype{C}[1]{>{\centering\arraybackslash}p{#1}}

\title{Infrared and Optical Emission-line Diagnostics}
\author[M. Durr\'{e}]{
Mark Durr\'{e}$^{1}$\thanks{E-mail: mdurre@swin.edu.au}
\\
$^{1}$Centre for Astrophysics \& Supercomputing, Swinburne University, Hawthorn VIC 3122, Australia\\
}

\date{Accepted XXX. Received YYY; in original form ZZZ}

\pubyear{2025}
\defcitealias{Stasinska2006}{[S06]}
\defcitealias{Kauffmann2003}{[KK]}

\begin{document}
\label{firstpage}
\pagerange{\pageref{firstpage}--\pageref{lastpage}}
\maketitle

\begin{abstract}
We study a catalogue of over 130 emission-line galaxies with matched near infra-red (NIR) and optical spectra, where we examine the relationship between the respective nuclear activity classifications, diagnosed by the flux ratios of emission lines. We match the standard NIR classification with four different optical classifications. While there is a broad agreement between the two regimes, there are mismatches and overlaps caused either by aperture effects and/or NIR radiation penetrating obscuring dust and ``seeing deeper'' into the nuclear region, thus exposing AGN activity. We examine the relationship between the equivalent widths {(EW)} of \Ha{} and \pab, as well as the ratios \NII/\Ha{} vs. \Fe/\pab, and find reasonable correlations. We thus propose a new diagnostic ({EW} of \pab{} with Fe - WPF) in the NIR (analogous to the WHaN classification), using the \Fe/\pab{} flux ratio and the {EW} of the \pab{} line. We show, within the limitations of the catalogue size, that the regions of the standard NIR diagram can be reasonably replicated in this new scheme. This diagnostic has the advantage that only one wavelength range needs to be observed, thus being economical with telescope time.
\end{abstract}




\section{Introduction}
The gas excitation in galaxies that show nuclear emission lines in their spectrum have several mechanisms. These are: photo-ionisation by star formation and active galactic nuclei (AGN) emissions \citep{Ho1993}, shocks from outflows and winds from AGNs and evolved stars \citep{Hollenbach1989}, direct heating of gas masses by X-ray emissions from AGN accretion disks \citep{Maloney1996} and UV fluorescence \citep[in the case of molecular hydrogen,][]{Black1987}. A galactic nucleus may have any combination of these excitation mechanisms, so it is crucial that the relative contributions can be characterised.

This is done by using diagnostic diagrams, where the ratios of various emission lines are plotted against each other. The locus of the object on the diagram shows the predominant excitation mechanism. These diagrams have been devised for both the optical and near infrared (NIR) electromagnetic regimes. This paper examines whether the diagnostics from these two regimes can be compared, using a survey of infrared and optical spectra from galaxy catalogues. We also introduce a new infrared diagnostic, comparable to the WHaN optical diagnostic of \cite{CidFernandes2010a}. 

This paper is organised as follows: the activity diagnostics are reviewed in \S\ref{sec:OIRD_2}, the spectral data and emission-line measurements are described in \S\ref{sec:OIRD_3}, in \S\ref{sec:OIRD_4} we discuss the results of the comparison of the emission-line diagnostics, as well as introducing the new diagnostic mentioned above, and we present our conclusions in \S\ref{sec:OIRD_5}.
\subsection{Activity Classifications}
\label{sec:OIRD_2}
\subsubsection{Optical Classification}
The two-dimensional optical classification scheme was originally proposed by \citet*{Baldwin1981} (the BPT diagram); this was enhanced by \cite{Veilleux1987} who introduced the standardized species flux ratios of {\NII($\lambda6583$)/\Ha{},  \SII($\lambda6716+6731$)/\Ha{} and \OI($\lambda6300$)/\Ha, all vs. \OIII($\lambda5007$)/\Hb}.  The advantage of these species ratios is that, apart from the lines being among the strongest in the optical, the wavelengths of each pair are close to each other and thus differential reddening and flux calibration uncertainties become inconsiderable factors. (In this work, we do not use the \OI/\Ha{} ratio, as the \OI{} flux is usually weak in comparison to the other species).

{\cite{Kewley2001a}, in their theoretical study of starburst galaxies, revised the \cite{Veilleux1987} separation between SF and AGN.} The advent of the Sloan Digital Sky Survey (SDDS) allowed much larger samples (many thousands of galaxies vs. a few hundred for previous studies). \cite{Kauffmann2003} examined over 22\,000 galaxies and derived a somewhat different demarcation between SF and AGN to \cite{Kewley2001a}; the region between the two demarcations was classified as ``Composite'', i.e. showing characteristics of both SF and AGN mechanisms. \cite{Kauffmann2003} also divided the AGN region into Seyfert and low-ionization nuclear emission-line region (LINER) components, using a position angle on the diagram determined by the \OIII{} luminosity ($\Phi \approx 25\degr$). {This line is defined as:~log(\OIII/\Hb)~=~1.01~log(\NII/\Ha)~+~0.48. This line is indistinguishable from that of \cite{Schawinski2007}: log(\OIII/\Hb)~=~1.05~log(\NII/\Ha)~+~0.45.} This study was extended by \cite{Kewley2006} with approximately 4 times the number of objects, further distinguishing Seyfert from LINER galaxies. Hereafter, the combined \cite{Kewley2001a} and \cite{Kauffmann2003} schemes are referred to as \citetalias{Kauffmann2003}. The \cite{Kewley2001a} and \cite{Kauffmann2003} classification based on \SII/\Ha{} is labelled \SII{} for the rest of this paper. 

\cite{Stasinska2006} (hereafter \citetalias{Stasinska2006}),  based on observation and theoretical photoionisation models, showed that the dividing line between SF and AGN for the BPT diagram is somewhat stricter than the \cite{Kauffmann2003} line (i.e. decreasing the size of the SF regime), excluding any mixed AGN component.

Observationally, those galaxies that have weak emission lines (e.g. the lines are absent or below the required signal to noise ratio) cannot be placed on the BPT diagram. Using SDSS with a $\sim$370\,000 galaxy sample, \cite{CidFernandes2010a,CidFernandes2011} introduced the WHaN diagnostic, i.e. replacing the \OIII($\lambda5007$)/\Hb{} with the equivalent width (EW) of \Ha{} (\wha). This allowed classification of many more galaxies, was economical with spectroscopic time (since only one spectral region need be observed), and also enabled further division of AGNs into strong (i.e. Seyferts) and weak (i.e. true LINERs powered by black hole accretion). These works also identified retired galaxies (``fake AGN''), where any black hole activity is weaker (or, at most, comparable) with ionization from hot, low-mass evolved stars {\citep[HOLMES]{Stasinska2008}. These can be either planetary nebula central stars, hot pre-white dwarfs which have lost their envelope and/or post-asymptotic giant branch (p-AGB) stars \citep{Yan2012},} as well as passive galaxies (i.e. very low or no activity). Emission from HOLMES and supernova remnants (SNRs) can also be non-nuclear, in which case they are sometimes labelled LIERs (e.g. low-ionization emission-line regions). This diagnostic has been used in subsequent studies to spatially distinguish between HOLMES and AGN photoionisation in IFU observations \citep[e.g.][]{Kehrig2012, Lopez-Coba2020, Heckler2022,Mezcua2024}, as well as incorporating it into galaxy censuses \citep[e.g.][]{Herpich2016}. 

{Some LINERs are certainly powered by BH accretion; the survey of local LINERs of \cite{Cazzoli2018} showed the universality of the presence of the broad-line region (BLR), strongly supporting the AGN mechanism. Similar, the sample of the X-ray data from 82 LINERs by \cite{Marquez2017} shows that 60--80\% of them can be considered genuine AGNs, that bright LINERs are similar to Seyfert 2 AGNs in their mid-IR spectroscopic properties (with faint LINERs being separate, thought to be from the disappearance of the dusty torus) and that HST imaging shows broad-cone or core-halo morphology, indicative of outflows.}  

However, the WHaN diagnostic has been critiqued \cite[e.g.][]{Torres-Papaqui2024}, as not allowing unambiguous identification of p-AGB vs. true AGN LINERs; studies deduced that the p-AGB integrated light in galaxies is generally low \citep[see][]{Conroy2013}. On the other hand, \cite{Nemer2024} have shown that LINERs can be identified by absorption lines in the continuum without using emission-line information. Using machine-learning classification on MaNGA integral field unit (IFU) spectra, they were able to identify LINER sources from the stellar continuum alone, and the deduced stellar population was consistent with evolved low mass stars. {See also \cite{Stasinska2025} for a review.}

Galaxies, of course, have multiple excitation sources; untangling these in the era of IFU spectroscopy show the limitations of these previous schemes based on single-slit spectroscopy. \cite{DAgostino2019a}, using NGC\,1068 as a test case, showed how the spatially resolved emission-line fluxes, coupled with gas kinematics (velocity dispersion) and physical radius from the nucleus, can be used to separate regions of pure star formation, AGNs, and shock ionization from one another. \cite{Johnston2023} further tested this on the SAMI survey, showing the general applicability of these techniques to a wide range of galaxy masses and morphological types. This study incorporated gas velocity dispersion on spatially resolved spectra from IFU observations, to distinguish shock components from photo-ionisation.

There are limitations to these schemes to diagnose excitation mechanisms. For example, X-ray bright (i.e. certainly containing an AGN) but otherwise optically ``dull'' with no or weak optical emission lines (X-ray Bright Optically Normal Galaxies - XBONGs), are shown by \cite{Agostino2023} to have genuinely weak emission that is not due to aperture effects (continuum swamping or star formation). The study by \cite{Herpich2016} uses the WHaW (EW of \Ha{} vs the \textit{WISE W2-W3} colour) to demonstrate that LINER-like galaxies are a mixture of different physical phenomena and observational effects, that all land on the same locus in the BPT diagram.

The \NII-\citetalias{Kauffmann2003} and the \SII{} classifications can give inconsistent types for the same object. \cite{Ji2020} have proposed a 3D re-projection of these diagrams, with the axes being the \NII/\Ha, \SII/\Ha{} and \OIII/\Hb{} ratios, putting strong constraints on the photo-ionization models and finding that nitrogen abundance prescriptions (relative to oxygen) must be different for SF and AGN activity. Their study also showed that composite objects can be decomposed into SF and AGN/LINER(LIER) regimes with well-constrained fractions of contamination from other sources. 

\subsubsection{NIR Classification}
Emission line objects can be categorized in the NIR by the diagnostic ratios \Fe/\pab{} and \Htwo/\brg{} introduced by \cite{Larkin1998}, with \cite{Rodriguez-Ardila2005} delineating the diagnostic diagram regions for star formation (SF), AGN and LINER excitation. The boundaries of these regions were updated by \cite{Riffel2013a} and included a class of ``transitions objects'' (TOs), that had high line ratios but were not classified as LINERs. {In this work, we follow \cite{Riffel2021a} and refer to transition objects, LINERs and supernovae remnants as ``high line ratio'' objects (HLR) as they all exhibit a wide range of excitation mechanisms, especially for \Htwo, as noted before.} 

Other diagnostics, based on photo-ionization models and incorporating other metallic species (\SIII{} 953 nm, \CI{} 985 nm, \PII{} 1188 nm and \Fe{} 1640 nm) linked to \pab{} and \pag~\citep{Calabro2023}, distinguish between SF- and AGN-driven photo-ionization in galaxies. These diagnostics are successfully applied up to high redshifts (\textit{z}$\sim$3), and have the further advantage that they do not require cryogenically cooled instruments for local (low \textit{z}) objects.
\section{Data and Measurements}
\label{sec:OIRD_3}
Our data consist of a matched set of optical and NIR spectral observations of galaxies. We start from the NIR spectral sample, confirm that they have emission lines and {can be placed on the diagnostic diagram}, then examine the corresponding optical spectrum for each object. 
\subsection{NIR Data}
The NIR sample consists of four surveys; (1) \cite{Riffel2006}, an atlas of AGN of all activity types, analysing continuum and emission line spectral properties, (2) \cite{Mason2015}, a NIR follow-up to the Palomar nearby galaxy survey \citep{Ho1995,Ho1997}, finding most of these galaxies contain AGN with a wide range of luminosities, as well as studying stellar populations, (3) \cite{Mould2012} and \cite{Durre2023}, a NIR spectral sample of early-type galaxies with radio emission, finding significant AGN and/or star formation activity, and (4) \cite{Martins2013}, an atlas of \HII{} (or starburst - SB) galaxies, which found no or poor correlation between optical and NIR spectral indices and indications that NIR emission lines were not generated in the nucleus. 

The characteristics of the samples are given in Table \ref{tbl:OIRD01}, which lists the instrument and telescope, the spectroscope mode (SXD - short cross-dispersed, XD - cross-dispersed), the slits size in arcsec, the apertures i.e. the length along the slit for data extraction, the total number of objects in each survey and the number with an adequate set of NIR emission lines.  Removing duplicates between catalogues, there are 275 objects. The total number of objects that can be classified onto the NIR diagnostic diagram is 132.
The spectral data were acquired either from the NASA/IPAC Extragalactic Database (NED)\footnote{\url{http://ned.ipac.caltech.edu/}} by downloading individual spectra, or from the published catalogues.
\begin{table*}
\centering
\caption{NIR catalogue details.}
\begin{tabular}{@{}clllllrrrr@{}}
\toprule
Catalogue & Reference                   & Telescope          & Instrument & Mode & Slit & FWHM&Objects & Good Lines & Data Source \\ 
&&&&(\arcsec)&(nm)&&&\\\midrule
1       & \cite{Riffel2006}           & IRTF               & Spex       & SXD  & 0.8 x 15       & 1.6&49         & 46            & 1           \\
2       & \cite{Mason2015}            & Gemini North       & GNIRS      & SXD  & 0.3 x 7        & 1&46         & 28            & 2           \\
3       & \cite{Mould2012} & Palomar 200\arcsec & TripleSpec & XD   & 1 x 30         & 0.65&152        & 49            & 3           \\
&\cite{Durre2023}&&&&&&&&\\
4       & \cite{Martins2013}          & IRTF               & Spex       & SXD  & 0.8 x 15       & 1.6&28         & 14            & 1  \\ \bottomrule        
\end{tabular}%
\begin{flushleft}
Data source - (1) NASA/IPAC Extragalactic Database (NED). (2) Canadian Advanced Network for Astronomical Research (CANFAR) \url{https://www.canfar.net/storage/vault/list/karun/xdgnirs_Dec2014} (3) \cite{Durre2023} \url{http://vizier.cds.unistra.fr/viz-bin/VizieR-3?-source=J/MNRAS/524/4923}
\end{flushleft}
\label{tbl:OIRD01}
\end{table*}

\subsection{Optical Data}
Of the 132 objects that have an NIR classification, optical data were retrieved from a variety of sources. Table \ref{tbl:OIRD02} lists the sources of the optical spectra, with the source reference, survey program and/or archive source, plus the number of objects. The majority of the spectra were retrieved from NED, with others from web archive sources. 5 objects without spectra had tabular data of the line fluxes or ratios, and two spectra were obtained by digitising plots from the references. 

3 objects had no optical spectrum or other data (2MASXJ20173144+7207257, Mrk\,0504 and Mrk\,0896) and one objects (UGC\,11228) had poor quality spectra, leaving 128 objects. 

\begin{table*}
\centering
\caption{Optical catalogue details.}
\begin{tabular}{@{}cllrl@{}}
\toprule
Catalogue & Reference                    & Program/Archive        & Objects & Data Source                                           \\ \midrule
1       & \cite{Jones2009}             & 6dF GS                 & 15      & NED                                                   \\
2       & \cite{Oh2022}                & BASS DR2               & 29      & \url{http://www.bass-survey.com/dr2.html}             \\
3       & \cite{Ho1995}                &                        & 42      & NED                                                   \\
4       & \cite{Abdurrouf2022}         & SDSS                   & 13     & \url{https://www.sdss4.org/dr17/}                     \\
5       & \cite{VanDenBosch2015}       & HETMGS                 & 3       & Tabular data                                          \\
6       &                              & CfA Archive            & 3       & \url{https://oirsa.cfa.harvard.edu/search/}           \\
7       & \cite{Moustakas2006}         &                        & 6       & NED                                                   \\
8       &                              & NOIRLab Archive        & 3       & \url{https://astroarchive.noirlab.edu/portal/search/} \\
9       & \cite{KennicuttRobertC.1992} &                        & 2       & NED                                                   \\
10      & \cite{Falco1999a}             & Updated Zwicky Catalogue & 7       & NED                                                   \\
11      & \cite{Ohyama1997}            &                        & 1       & \url{https://smoka.nao.ac.jp/fssearch}                \\
12      & \cite{Shuder1981}            &                        & 1       & Tabular data                                          \\
13      & \cite{Kewley2001}            &                        & 1       & Digitised graph                                       \\
14      & \cite{Moran2000}             &                        & 1       & Digitised graph                                       \\
15      & \cite{Menezes2022}           & DIVING$^{3D}$          & 1       & Tabular data                                          \\   
\bottomrule
\end{tabular}%
\label{tbl:OIRD02}
\end{table*}
\subsection{Emission Line Flux Measurements}
For the NIR spectra, the emission lines are \Fe~(1257 nm), \pab~(1282 nm), \Htwo~(2121 nm) and \brg~(2166 nm). For the optical spectra, the emission lines are \Ha~(6564 \AA), \Hb~(4863 \AA), \OIII~(5007 \AA), \NII~(6584 \AA) and the doublet \SII~(6717 and 6731 \AA).  We fit a  Gaussian function to the forbidden (\Fe, \NII, \OIII~and \SII) and molecular hydrogen \Htwo{} emission lines. {Only the narrow-line component of the permitted hydrogen lines (\pab, \brg, \Ha{} and \Hb) is used in the activity classifications. The number of fitting components (especially for the hydrogen lines) is chosen by examining the shape of the forbidden lines, i.e. whether line wings or asymmetries (caused by outflows) are visible. Symmetrical wings can be fitted with a single Lorentzian function (or two Gaussians), whereas asymmetries can be fitted with two Gaussian components. If the hydrogen lines have a significantly different width (or shape), then broad-line components have to be added. (See section \ref{sec:OIRD_34})}

{The galaxy spectra are first fitted for the stellar continuum using the pPXF routine \citep{Cappellari2017}, using a selection of eMILES single stellar population (SSP) templates \citep{Vazdekis2016} with 6 metallicities ([Z/H] in the range [-1.71, 0.22]) and 25 ages (with log ages in Myr from 1.8 to 4.2 in steps of 0.1). When fitting the infrared spectra, we exclude ages less than 300 Myr, as these templates are not computed in the SSP set at these wavelengths. This is not an issue, as the stellar light is completely dominated by older populations in this wavelength range. The procedure logarithmically rebins and normalises the galaxy spectrum, then masks the set of prominent emission lines over the wavelength range. The pPXF routine then fits the galaxy spectrum to the templates, including multiplicative polynomials to adjust the continuum shape of the template. The routine returns the best fit model spectrum; this is subtracted from the original spectrum to remove the stellar continuum, leaving just the emission lines.}

We then use the \texttt{QFitsView}\footnote{\url{http://www.mpe.mpg.de/~ott/dpuser/qfitsview.html}} \citep{Ott2012} ``de-blending'' functionality to fit both NIR and optical emission lines; this allows for both Gaussian and Lorentzian function fits with single or multiple components. This requires some manual input from the user to set the initial estimates of continuum, height, centre and {full width half-maximum (FWHM)} of each component. It  uses the GSL\footnote{\url{https://www.gnu.org/software/gsl/doc/html/index.html}} \textit{gsl\_multifit} routines, returning fit values and errors of each component (continuum slope, central wavelength, FWHM and flux). 

\begin{figure*}
    \centering
    \includegraphics[width=1\linewidth]{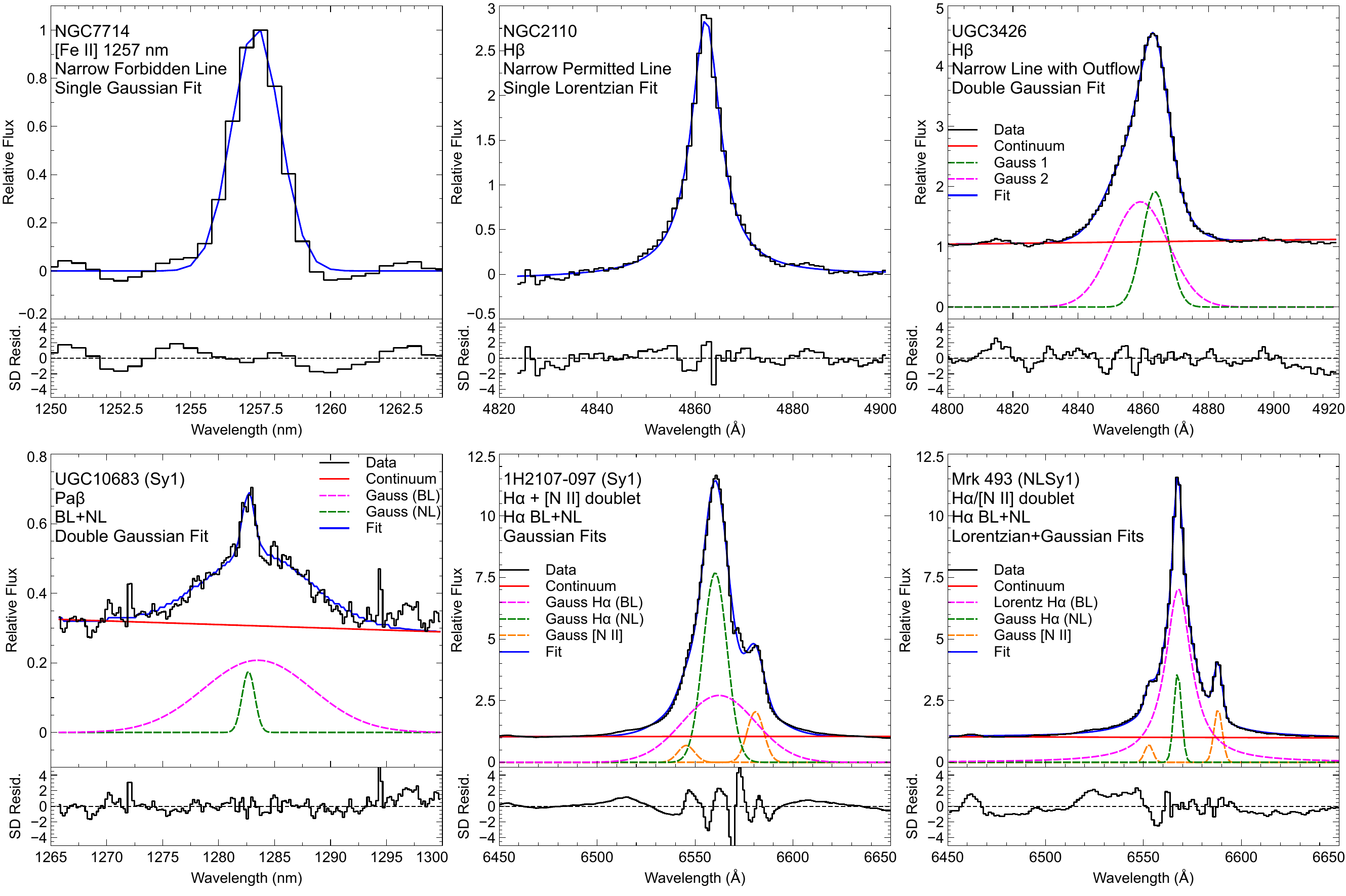}
    \caption{Spectral line fitting examples. {Top left: single Gaussian fit. Top middle: Narrow permitted line modelled with a Lorentzian fit. Top right: Narrow-line outflow modelled with 2 Gaussians. Bottom left: Broad and narrow line fit for a Seyfert 1. Bottom middle: fit for the \Ha-\NII{} line complex, with broad and narrow \Ha{} components and the two narrow, forbidden \NII{} lines, all modelled with Gaussian fits. Bottom right: as previously, but with a Lorentzian fit to the broad \Ha{} line for a NLSy1 type. The black lines are the original spectral data, the blue line is the total fit to the data, the red dashed lines are the broad line components and the green and orange dashed lines are the narrow line components. The residuals (Data - Fit)/$\sigma$(Data - Fit) are plotted below each panel. }All spectra have been flux normalised.}
    \label{fig:WPFPlots08}
\end{figure*}

\begin{figure*}
    \centering
    \includegraphics[width=1\linewidth]{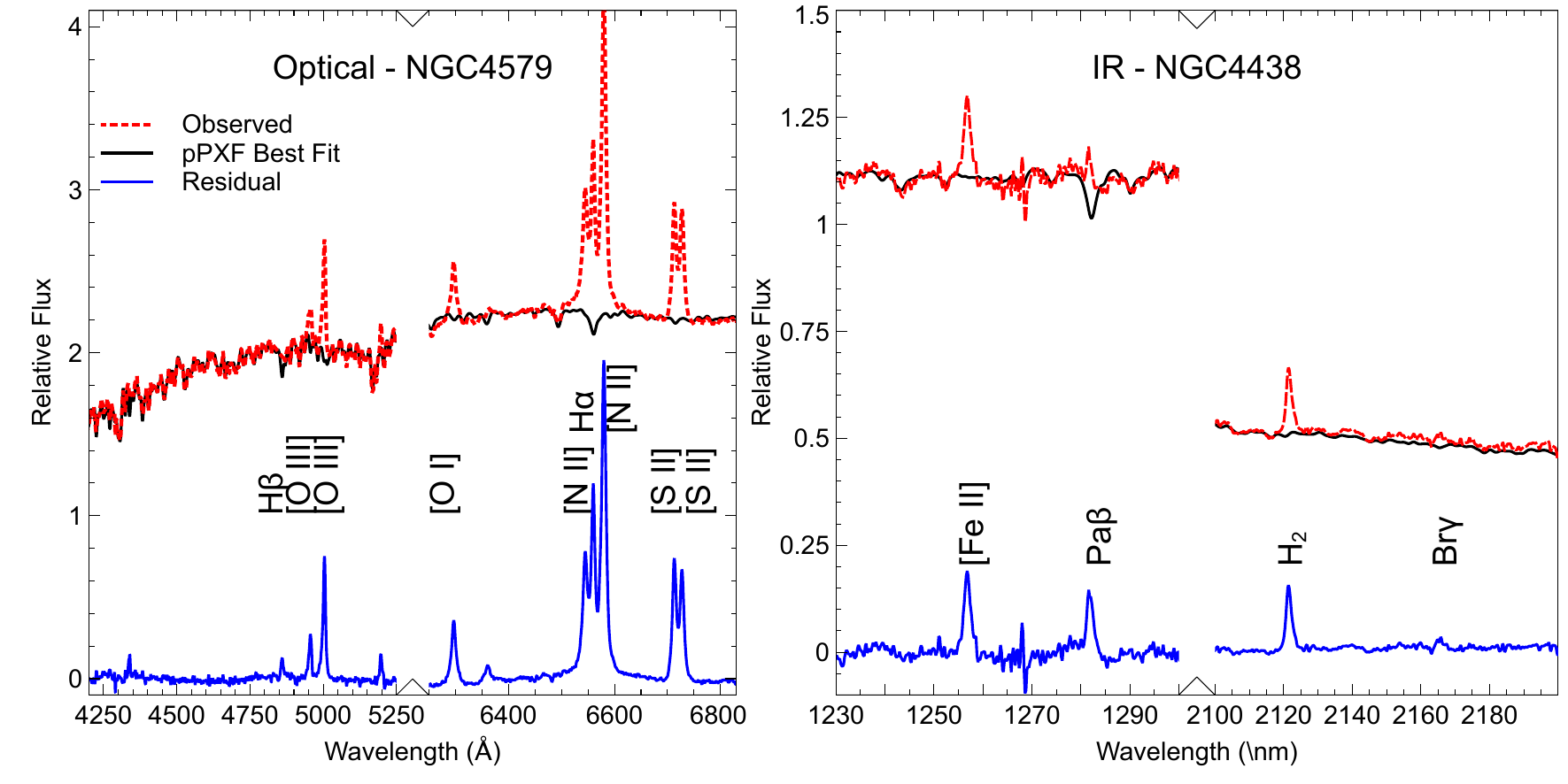}
    \caption{Example pPXF spectral fitting. Left panel: optical spectrum fitting for NGC\,4579. Right panel: IR spectral fitting for NGC\,4438. This shows the original spectra (red), the best fits from the pPXF routine (black) and the resulting residuals (blue). Note that the flux of the H lines (\pab, \brg, \Ha{} and \Hb) changes relative to the narrow lines. The spectra have been flux normalised.}
    \label{fig:WPFPlots11}
\end{figure*}

Example fits are given in Figure \ref{fig:WPFPlots08}, showing a single Gaussian fit, a combination of broad and narrow permitted lines and an example of fitting the \Ha-\NII{} complex with broad and narrow \Ha{} lines and the two forbidden \NII{} lines. {An example pPXF full-spectrum fit of NGC\,4579 is also plotted (Figure \ref{fig:WPFPlots11}), showing the best fit and residuals; this plot illustrates the difference in flux ratios, resulting from the increased flux of the H lines. Without the pPXF fit, the flux ratios are log(\NII/\Ha) = 0.32 and log(\OIII/\Hb) = 0.58 (putting it in the LINER regime); the residuals show the respective ratios as 0.25 and 0.24, i.e. the location on the BPT diagram is closer to the SF regime.}

For NIR spectra, both the \Fe{} and \Htwo{} lines must be present;  if both hydrogen lines are missing, the upper limit of the flux is estimated from the 3$\sigma$ noise over a 10 nm window around the line wavelength.  In every case where the \pab{} line is not visible, the \brg{} line is also not seen. If the \brg{} line is not present, then the value is calculated using the ratio \brg/\pab{}~=~0.171, which is the canonical ratio for case B recombination (assuming an electron temperature $T_e = 10^4$ K and a density $n_e=10^3~cm^{-3}$) \citep{Hummer1987}. 10\% (14/132) of the NIR spectra use this ratio.

For the optical measurements, the \SII{} doublet was not measured for several reasons; tabular data is not available (NGC\,0708, NGC\,2629 and NGC\,7013), the line is out of the spectral range (Mrk\,573), or was too noisy to measure (NGC\,2706 and NGC\,7648). One object (NGC\,2342) had \OIII $\lambda5007$ out of the spectral range, so the value was calculated from the  \OIII $\lambda4959$ flux, with the ratio \OIII{}$\lambda$5007/4959~=~2.98 \citep{Storey2000}.

\Hb{} was either not present or was to noisy to measure accurately for 5 objects. In these cases, we used the same method as for the missing \brg{} line in the NIR spectra; in this case using the ratio \Hb/\Ha{}~=~0.35. 
Two objects had neither \OIII{} or \Hb{} measurements, due to noisy or missing spectra; these were included for the WHaN measurement.

\subsection{Broad-Line Components}
\label{sec:OIRD_34} 
For some AGN (type 1), the spectrum shows large line widths for the permitted lines (mainly H, He and O species); these come from the broad-line region (BLR), \rev{where high-velocity gas clouds are close to the SMBH}, and have typical FWHM velocities $>$~1000 ~\kms. In type 2 AGN, these are not present, presumably hidden from the line-of-sight by obscuring material. The presence of broad lines is deduced by the line width being significantly greater than for the forbidden lines (e.g. \OIII{} or \Fe), or by obviously symmetric line wings.  The broad line is often not visible in \Hb{} (Seyfert 1.9 types) \rev{but it is seen} in \Ha{} - this is thought to be mainly from differential dust extinction, usually when the broad \Ha{} line is weak; similarly this is also the case for the \brg{} vs. the \pab{} lines, though this is more due to the intrinsic line strength ratio.

For the optical spectra, 23 objects (of 128) had broad hydrogen lines, with 7 objects preferring a Lorentzian fit to the broad line. For the infrared spectra, 32 objects (of 132) have broad lines, with 7 using Lorentzian fits. {5 other objects showed signs of outflow, and were fitted with 2 Gaussians; in these cases, the flux for each is summed.} 2 objects, NGC\,4151 and NGC\,5548, used 2 and 3 BL components, respectively, to make a good fit. 8 objects have broad NIR lines without broad optical lines; this is most likely due to NIR radiation penetrating obscuring dust and ``seeing deeper'' into the AGN core. One object, NGC\,3147, has optical broad lines with no NIR lines; this is due to a poor quality NIR spectrum. 

Fitting the \Ha-\NII{} complex is problematic in some cases, because the lines can be severely blended. {For these spectra, we fitted a combination of Gaussians to the spectral region around the complex, using the generalized reduced gradient algorithm (`GRG Nonlinear') implemented in the \texttt{MS-Excel} add-on \textit{Solver}. The \Ha{} emission is modelled with a combination broad and narrow Gaussians, plus two \NII{} Gaussians, whose central wavelengths are fixed with respect to the narrow \Ha{} Gaussian and the flux ratio \NII~$\lambda 6583/\lambda 6549 = 3$  \citep{Osterbrock2006}. Some authors \citep[e.g.][]{Schmidt2016} find a requirement for an `intermediate component' contributing to the broad-line profile. In our sample, there was no case where this was needed to provide a good fit.} 

\begin{figure}
    \centering
    \includegraphics[width=1\linewidth]{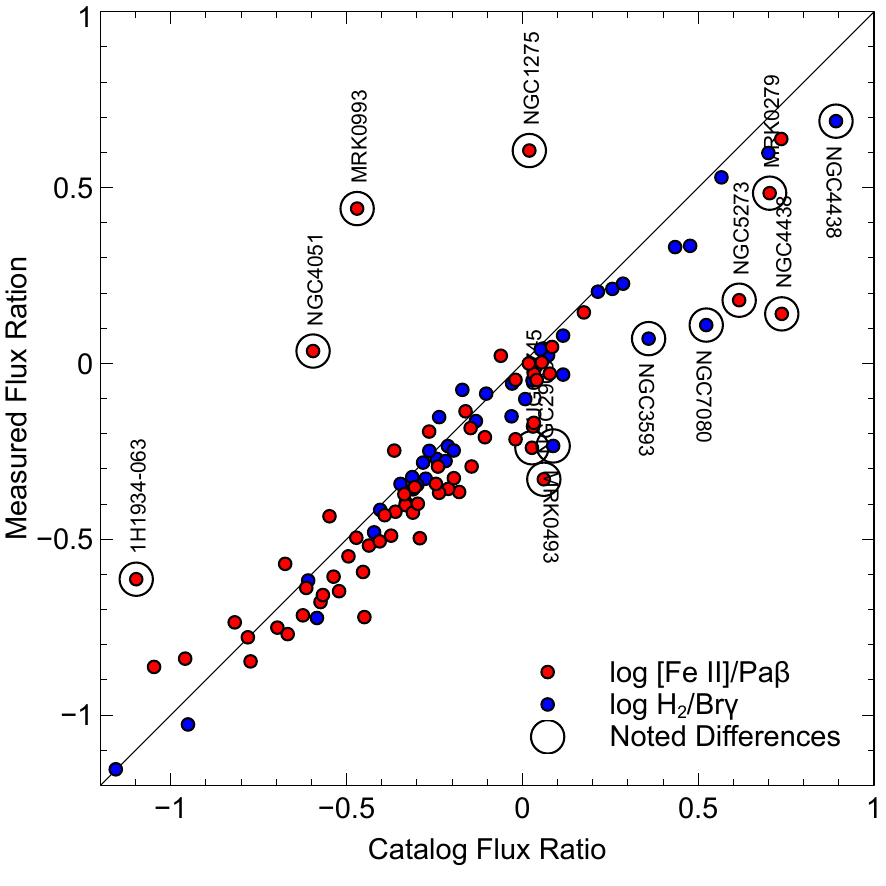}
    \caption{{Log flux ratio comparision for \Fe/\brg{} (red points) and \Htwo/\pab{} (blue points). Points that lie off the 1:1 ratio are circled; see the text for details.}
    }
\label{fig:WPFPlots13}
\end{figure}

{A special case exists for Narrow-Line Seyfert 1 (NLSy1) galaxies, of which our catalogue contains 10 objects. The profile of the broad lines is considered to be best fitted with a Lorentzian curve; this is physically motivated, both on observational \citep{Kollatschny2013, Berton2020,Durre2021} and theoretical grounds \citep{Goad2012}. In these cases, we compared Gaussian and Lorentzian broad-line fits using the F-test; the NLSy1 galaxies 1H1934-067, Ark\,564, Mrk\,124, Mrk\,291, Mrk\,478, Mrk\,493, Mrk\,1239, Mrk\,896 and NGC\,4051 preferred a Lorentzian fit. It is noted that a Lorentzian fit to the broad-line component will reduce the narrow-line \Ha{} flux somewhat, as it is more peaked than the Gaussian curve. Figure \ref{fig:WPFPlots08} shows examples of these broad-line fits.}
{\subsection{Comparisons with Catalogues}
We can compare the IR measured data from that given in tabular form by 3 of the 4 source catalogues \citep{Durre2023,Riffel2006,Martins2013}. \cite{Mason2015} does not provide this data. Figure \ref{fig:WPFPlots13} shows the comparison of the log ratios of \Fe/\brg{} (red points) and \Htwo/\pab{} (blue points); as can be seen, the values lie close to the 1:1 ratio line. The exceptions (circled points) result from several causes; (1) fitting and subsequent subtraction of the stellar population using pPXF with removal of the \pab{} and \brg{} absorption components  (e.g. NGC\,4438), (2) uncertainties with fitting the broad H lines with multiple components, especially when the narrow-line components are significantly weaker that the broad-lines ones, (3) some measurement issues with the original catalogue data, revealed when remeasuring the fluxes and (4) noisy or low-quality spectra, which increases the uncertainties, especially when the fluxes of the line pairs are similar.}
\section{Results and Discussion}
\label{sec:OIRD_4}
\subsection{NIR and Optical Diagnostic Diagrams}
\begin{figure*}
\centering
\includegraphics[width=1\linewidth]{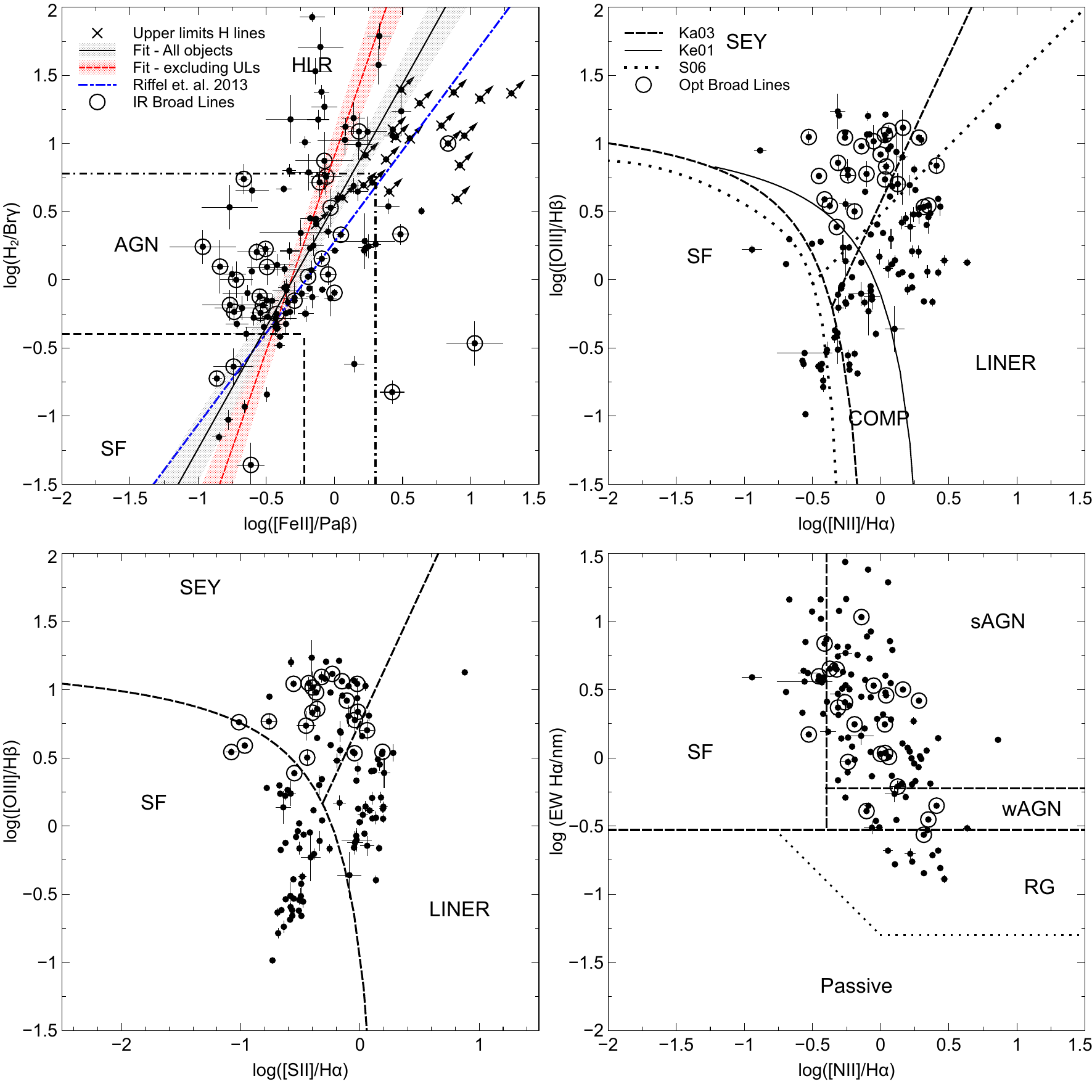}
\caption{{Top left: NIR diagnostic diagram. Objects with broad emission lines are designated with circles at each point. Objects with upper limits to the H lines are designated with arrows. The linear regression fits, including (solid black line) and excluding (dashed red line) HLR(UL) objects, show minimal difference to each other. The fit derived from \citet{Riffel2013a} is also plotted (dot-dashed blue line). Top right: optical \NII-\citetalias{Kauffmann2003} diagram. The solid line plots the \citet{Kewley2001a} [Ke01] demarcation with \citet{Kauffmann2003} [Ka03] regions delineate by dashed lines. The \citetalias{Stasinska2006} regions are delineated with dotted lines. Bottom left: \SII~ diagnostic diagram, showing greater proportion of SF objects. Bottom right: WHaN diagnostic diagram; the preponderance of objects are in the strong and weak AGN regimes.}
    }
    \label{fig:WPFPlots07}
\end{figure*}
Figure \ref{fig:WPFPlots07} plots the diagnostic ratios for the NIR (top left panel), optical \NII-\citetalias{Kauffmann2003} (top right panel), optical \SII{} (bottom left) and WHaN (bottom right) schemes. For the NIR, there is a good distribution of objects across the diagram. The objects with upper limits to the hydrogen line flux are plotted with limit arrows; in all cases where the \pab{} flux is an upper limit, so is the \brg{} flux. Almost 45\% of the high line ratio objects (21/48) have these upper limits; these will be denoted by HLR(UL), i.e. high line ratios with upper limits. We can say that these upper limits are realistic as they follow a smooth continuation from those points that have firm H flux measurements. 

We plot the orthogonal least squares regression for the NIR objects. This analysis (both for this figure and subsequent ones in this work) uses the BCES method of \cite{Akritas1996}\footnote{The BCES routine was from the Python module written by Rodrigo Nemmen \citep{Nemmen2012}, which is available at \url{https://github.com/rsnemmen/BCES}}; this regression method is preferred because there is no independent variable and both data sets have errors. We plot the 95\% confidence interval for the fit for all objects, which is:
\begin{equation}
\begin{split}
     \log(\rm{H_{2}/Br\gamma}) = (1.17\pm0.07)\log(\rm{[Fe II]/Pa\beta}) + (0.32\pm0.04)
\end{split}
\end{equation}
The regression line for objects excluding those with upper limits is:
\begin{equation}
\begin{split}
   \log(\rm{H_{2}/Br\gamma}) = (2.86\pm0.36)\log(\rm{[Fe II]/Pa\beta}) + (0.91\pm0.03)
\end{split}
\end{equation}
The difference between the slopes is mainly due to the group of HLR(UL) points in the top-right corner of the plot. The fit from \cite{Riffel2013a} is also plotted (solid red line). Note that the axes for the diagnostic plot are swapped from the convention from other works, i.e. the x-axis is the \Fe/\pab{} ratio. This is because we will use this as the x-axis later in this work. 

We also mark objects that have broad permitted lines with circles. It is noted that 3 objects in the SF region in the IR diagrams have broad lines (1H11934-063, Mrk\,504 and Ark\,564). In each case, the measurement uncertainties in the H lines are large due to blending, where the broad lines overwhelm the narrow lines. There is also a degree of dependence on the model fitting, with choices of Gaussian and/or Lorentzian fits giving different flux values.

The optical \NII-\citetalias{Kauffmann2003} plot shows a reasonable distribution of objects, with, perhaps, a deficiency of SF galaxies outside of the COMP region. The \cite{Kewley2001a} [Ke01] and \cite{Kauffmann2003} [Ka03] demarcations are plotted as solid and dashed black lines, respectively, with the \citetalias{Stasinska2006} regions delineated with dotted lines. Again, the broad-line objects are plotted with circles, with most being in the Seyfert region. The \SII{} plot places more galaxies in the SF region, with almost all of the broad-line galaxies being in the Seyfert region. {The WHaN diagram places preponderance of objects in the strong and weak AGN regimes.}
\subsection{Comparison of  NIR and Optical Types}
\begin{figure*}
    \centering
    \includegraphics[width=1\linewidth]{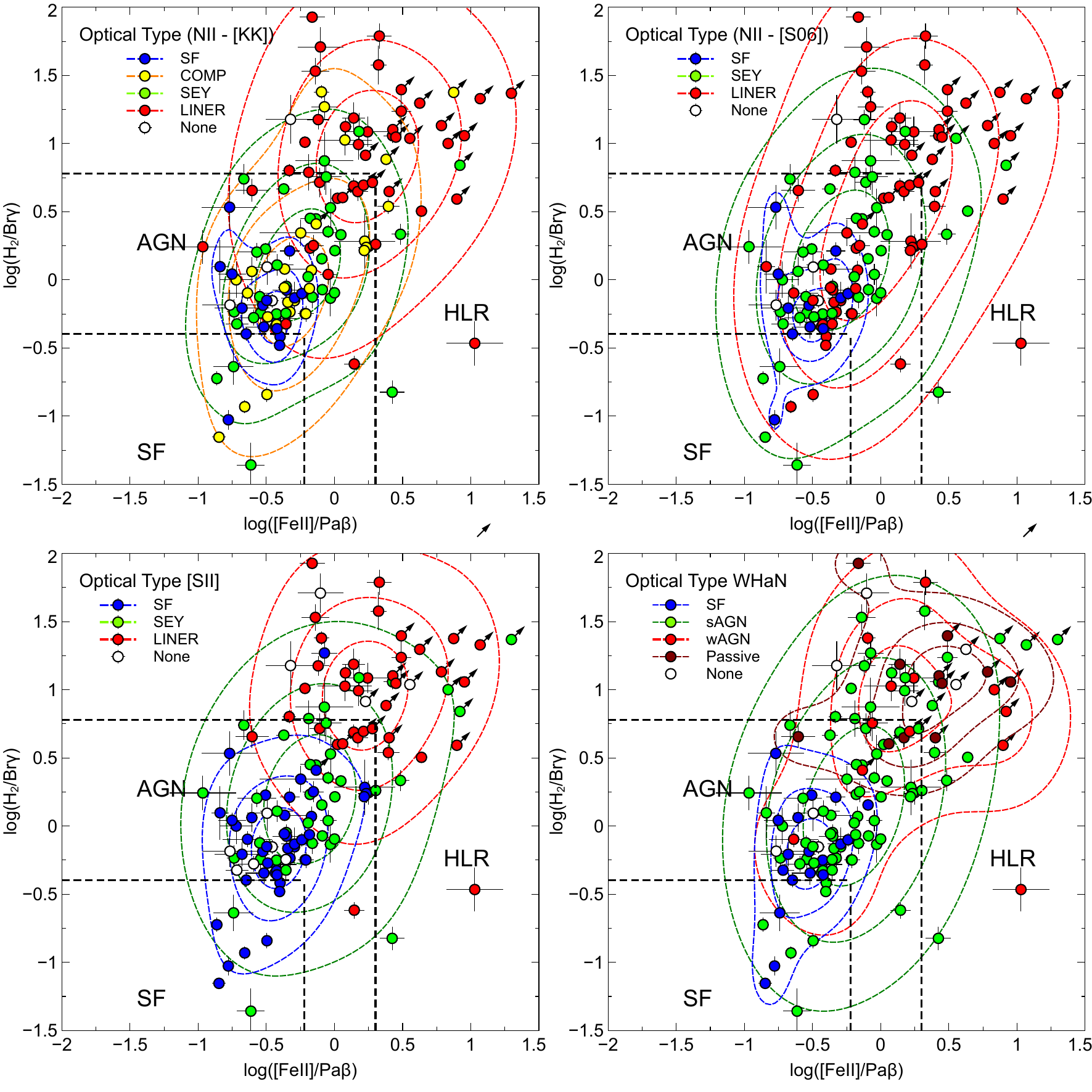}
    \caption{NIR classification \protect\citep[from][]{Larkin1998, Rodriguez-Ardila2005, Riffel2013a,Riffel2021a} colour-coded by optical classification. Top left: \protect\NII-\citetalias{Kauffmann2003} \NII/\Ha{} classification. Top right: \NII-\citetalias{Stasinska2006}  \NII/\Ha{} classification. Bottom left: \SII/\Ha{} classification. Bottom right: WHaN classification. The PDFs, calculated from Gaussian KDEs for each types, are plotted as contours, showing the 20, 50 and 80\% of maximum levels. Points with upper limits for \pab{} and \brg{} are shown with arrows.}
    \label{fig:WPFPlots01}
\end{figure*}

We can compare the classification of objects on the respective diagnostic diagrams, i.e. plot the NIR type on the optical diagram(s) and vice-versa. Figure \ref{fig:WPFPlots01} displays the optical classifications; the left panel shows the \NII-\citetalias{Kauffmann2003} classification, split by SF (blue), COMP - composite (yellow), SEY (green) and LINERs (red); the middle shows the \NII-\citetalias{Stasinska2006} scheme and the right panel plots the \SII{} classification. A few points have no optical classifications, as the available optical spectrum did not show any (or enough) emission lines. A majority (45/58) of optical LINER objects only have upper limits in the NIR for the hydrogen lines, again shown with limit arrows. 

We also plot a probability distribution function using a Gaussian kernel density estimate (KDE) for each optical type. The kernel bandwidth is estimated using Scott's rule \citep{Scott1992} {, which is $B = n ^ {-1/(d+4)}$ where $n$ is the number of data points and $d$ is the number of dimensions (in this case = 2). The KDE is then constructed as the sum of a set of 2D Gaussian curves, where $B \times \sigma_{x,y}$ is used as the Gaussian standard deviations, with $\sigma_{x,y}$ being the data standard deviations for the plot axes.} Contours are plotted at 20, 50 and 80\% of the peak value. 

As we can see, in all cases the optical SF points and KDEs are shifted more towards the AGN region with, in fact, the peak of the KDE actually in the AGN regions; showing a large degree of overlap between the two types. These can be understood as the NIR wavelengths penetrating deeper through nuclear dust to reveal the AGN; these are known as ``hidden'' AGN. This phenomenon has been noted in other studies \citep{Calabro2023, Lamperti2017, Onori2017}. {We can check that this is not due to imperfect de-blending of H lines, measuring incorrect narrow-line flux values; of the 15 objects that have a SF optical type and also have an AGN NIR type, only 2 have broad lines in the NIR, i.e. the rest are Seyfert 2 types.} Alternately, nuclear or peri-nuclear LIER or LINER emission may contribute, again obscured in the optical by dust. {All of these, of course, can be have aperture effects when comparing observations.}

For the \NII-\citetalias{Kauffmann2003} classification, it is noted that the COMP type seems to be completely subsumed by the SF and AGN types. This is confirmed by examining the distributions where the COMP type is classified either as SF or AGN; the KDEs are indistinguishable. That is, the COMP classification is not apparent in the NIR. It is also noted that the SEY and LINER classifications are cleanly separated, thus validating the demarcation between Seyferts and LINERs as introduced in \cite{Kauffmann2003}, which is not commonly used in the \NII-\citetalias{Kauffmann2003} scheme.

The \NII-\citetalias{Stasinska2006} classification shows a large degree of overlap between the AGN and LINER types; this can be viewed as due to the considerably stricter region delineations compared to \NII-\citetalias{Kauffmann2003}. The \SII{} classification has the cleanest separation between the types, though still with the large overlap of SF and AGN types.

\begin{figure*}
    \centering
    \includegraphics[width=1\linewidth]{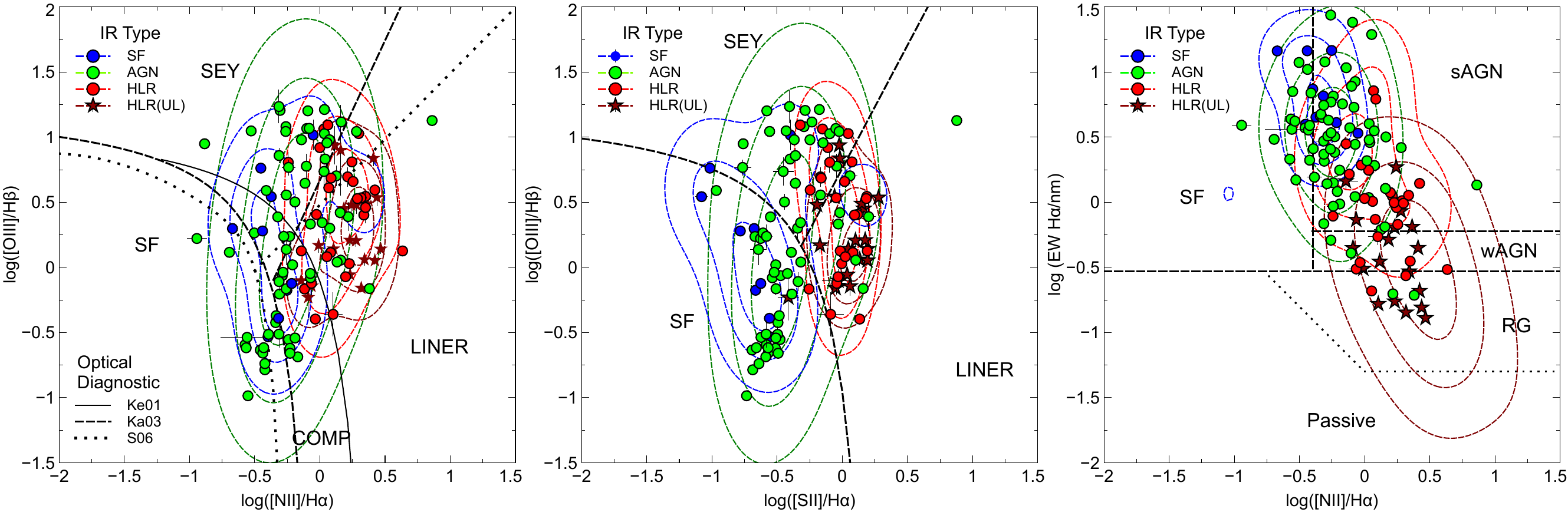}
    \caption{Optical classification, colour-coded by NIR classification. Left panel: \NII/\Ha{} classifications. Middle panel: \SII/\Ha{} classification. Right panel: WHaN classification}
    \label{fig:WPFPlots02}
\end{figure*}

In a similar manner, we can plot the optical classifications coded by the NIR types. Figure \ref{fig:WPFPlots02} shows the \NII/\Ha{} plot with the \NII-\citetalias{Kauffmann2003} and \NII-\citetalias{Stasinska2006} regions delineated (left panel), the \SII/\Ha{} plot (middle panel) and the WHaN plot (right panel). In all cases, there is a reasonably large degree of overlap between the NIR types. For the \NII/\Ha{} and \SII/\Ha{} plots, the NIR AGN types have significant numbers in the SF regime; this is the ``hidden'' AGN phenomenon as noted before. NIR HLR and HLR(UL) objects are firmly located in the LINER region of the \NII/\Ha{} plot with some overlap of HLR objects into the Seyfert region of the \SII/\Ha{} plot. The WHaN diagram has most NIR AGN and HLR types in the ``strong'' AGN region, with the locus of the HLR(UL) types mostly in the ``weak'' AGN and retired galaxies (RG) regions. The NIR SF type seem to overlap the AGN type in all diagrams; this may be due to the low number of objects (9). We also include the classification matching in tabular form, showing the percentages of NIR classification for each optical type (Tables \ref{tbl:OIRD03},\ref{tbl:OIRD04}, \ref{tbl:OIRD05} and \ref{tbl:OIRD010}).
\begin{table}
\centering
\caption{Classification matching for \NII-\citetalias{Kauffmann2003} vs. NIR diagrams The rows are the optical types and columns are NIR types. Numbers are percentages for each row. The last two rows assign the COMP classification to either SF or AGN types, showing almost identical distribution, i.e. the COMP type is not distinguishable in the NIR diagram.}
\label{tbl:OIRD03}
\begin{tabular}{lrrrr}
\toprule
&\multicolumn{4}{c}{NIR Type}\\
\NII-\citetalias{Kauffmann2003} Type& SF   & AGN  & HLR  & HLR(UL) \\ \midrule
SF & 16.7 & 83.3 & 0.0 & 0.0 \\
COMP & 10.7 & 64.3 & 14.3 & 10.7 \\
SEY & 8.8 & 76.5 & 11.8 & 2.9 \\
LIN & 2.1 & 27.1 & 41.7 & 29.2 \\ \midrule
SF+COMP & 13.0 & 71.7 & 8.7 & 6.5 \\
COMP+SEY & 9.7 & 71.0 & 12.9 & 6.5 \\\bottomrule
\end{tabular}
\end{table}
\begin{table}
\centering
\caption{As for Table \ref{tbl:OIRD03}, but for the \NII-\citetalias{Stasinska2006} classification.}
\label{tbl:OIRD04}
\begin{tabular}{lrrrr}
\toprule
&\multicolumn{4}{c}{NIR Type}\\
\NII-\citetalias{Stasinska2006} Type& SF   & AGN  & HLR  & HLR(UL) \\ \midrule
SF & 8.3 & 91.7 & 0.0 & 0.0 \\
SEY & 8.7 & 69.6 & 17.4 & 4.3 \\
LIN & 7.1 & 40.0 & 25.7 & 27.1 \\\bottomrule
\end{tabular}
\end{table}
\begin{table}
\centering
\caption{As for Table \ref{tbl:OIRD03}, but for the \SII{} classification.}
\label{tbl:OIRD05}
\begin{tabular}{lrrrr}
\toprule
&\multicolumn{4}{c}{NIR Type}\\
\SII{} Type& SF  & AGN  & HLR  & HLR(UL) \\ \midrule
SF & 15.9 & 79.5 & 2.3 & 2.3 \\
SEY & 5.1 & 66.7 & 20.5 & 7.7 \\
LIN & 2.6 & 18.4 & 39.5 & 39.5 \\ \bottomrule
\end{tabular}
\end{table}
\begin{table}
\centering
\caption{{As for Table \ref{tbl:OIRD03}, but for the WHaN classification.}}
\label{tbl:OIRD010}
{\begin{tabular}{lrrrr}
\toprule
&\multicolumn{4}{c}{NIR Type}\\
WHaN Type& SF   & AGN  & HLR  & HLR(UL) \\ \midrule
SF & 15.8 & 84.2 & 0.0 & 0.0 \\
sAGN & 7.3 & 62.2 & 23.2 & 7.3 \\
wAGN & 8.3 & 16.7 & 33.3 & 41.7 \\
PSV & 0.0 & 18.2 & 18.2 & 63.6 \\\bottomrule
\end{tabular}}
\end{table}
\subsection{\texorpdfstring{\pab{} and \Ha{}}{PabHa} Equivalent Widths}
\begin{figure*}
    \centering
    \includegraphics[width=1\linewidth]{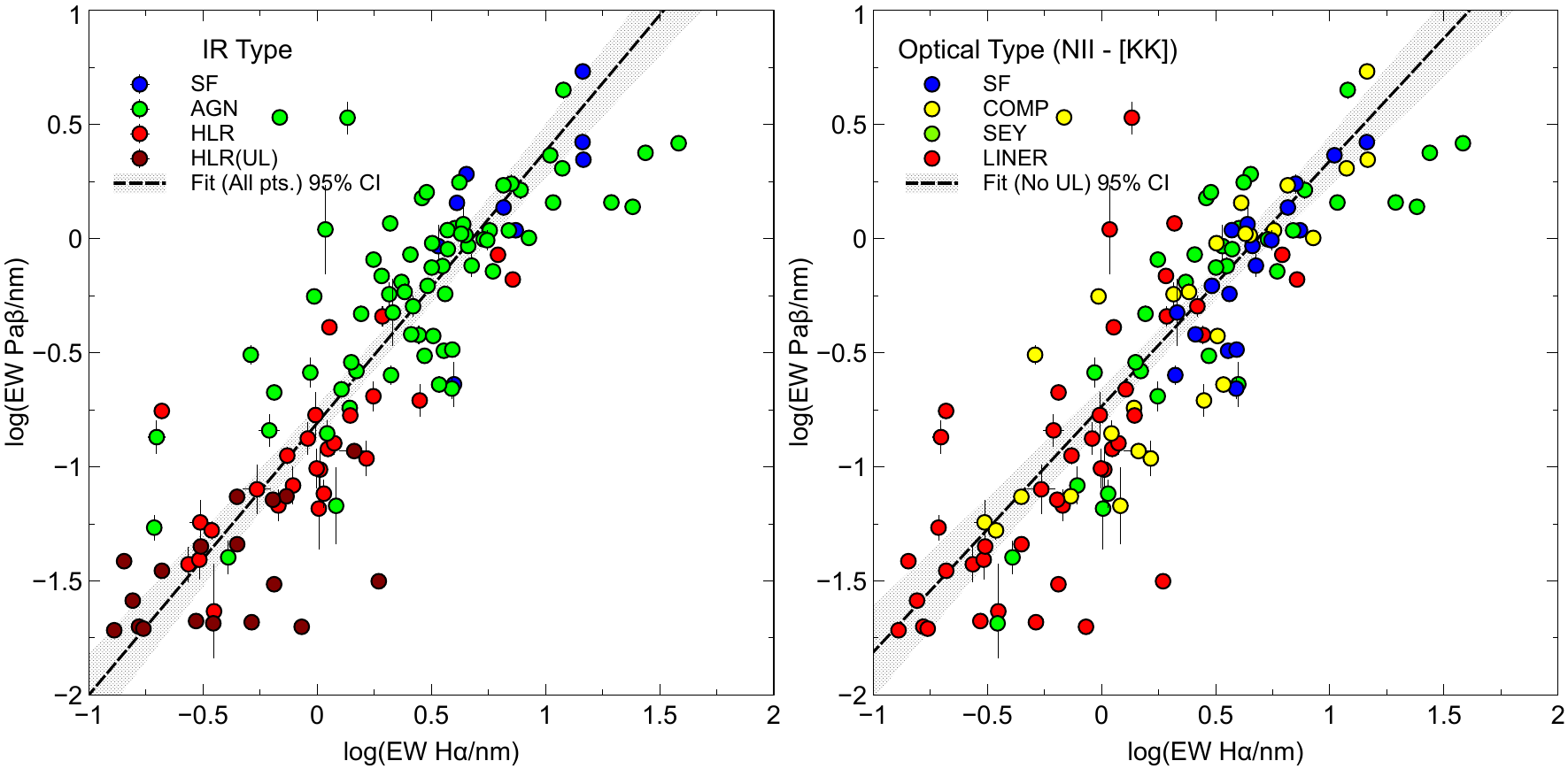}
    \caption{Plot of \wha{} vs. \wpb, colour-coded by NIR type (left panel) and \NII-\citetalias{Kauffmann2003} optical type (right panel), with linear regression fits with 95\% confidence intervals.}
    \label{fig:WPFPlots04}
\end{figure*}

By analogy with the WHaN diagnostic (\wha{} vs. \NII/\Ha) \citep{CidFernandes2010a,CidFernandes2011}, we examine whether \wpb{} can be used as a diagnostic. Figure \ref{fig:WPFPlots04} plots \wha{} vs \wpb, colour-coded by NIR and optical \NII{} types. The orthogonal least squares regression is also plotted with 95\% confidence intervals, with the fit:
\begin{equation}
    \log(W_{Pa\beta}) = (1.19\pm0.07)\log(W_{H\alpha}) - (0.81\pm0.04)
\end{equation}

The slope of the fit is close to 1; however the correlation is not strong ($R^{2} = 0.66$). The values with $\log(W_{Pa\beta})~<~-1$ show somewhat increased scatter around the regression line, due to these points only having upper limits to the \pab{} flux measurements. If we exclude these points, the slope is identical to 1 (within errors):
\begin{equation}
    \log(W_{Pa\beta}) = (1.08\pm0.08)\log(W_{H\alpha}) - (0.74\pm0.04)
\end{equation}

\begin{figure*}
    \centering
    \includegraphics[width=1\linewidth]{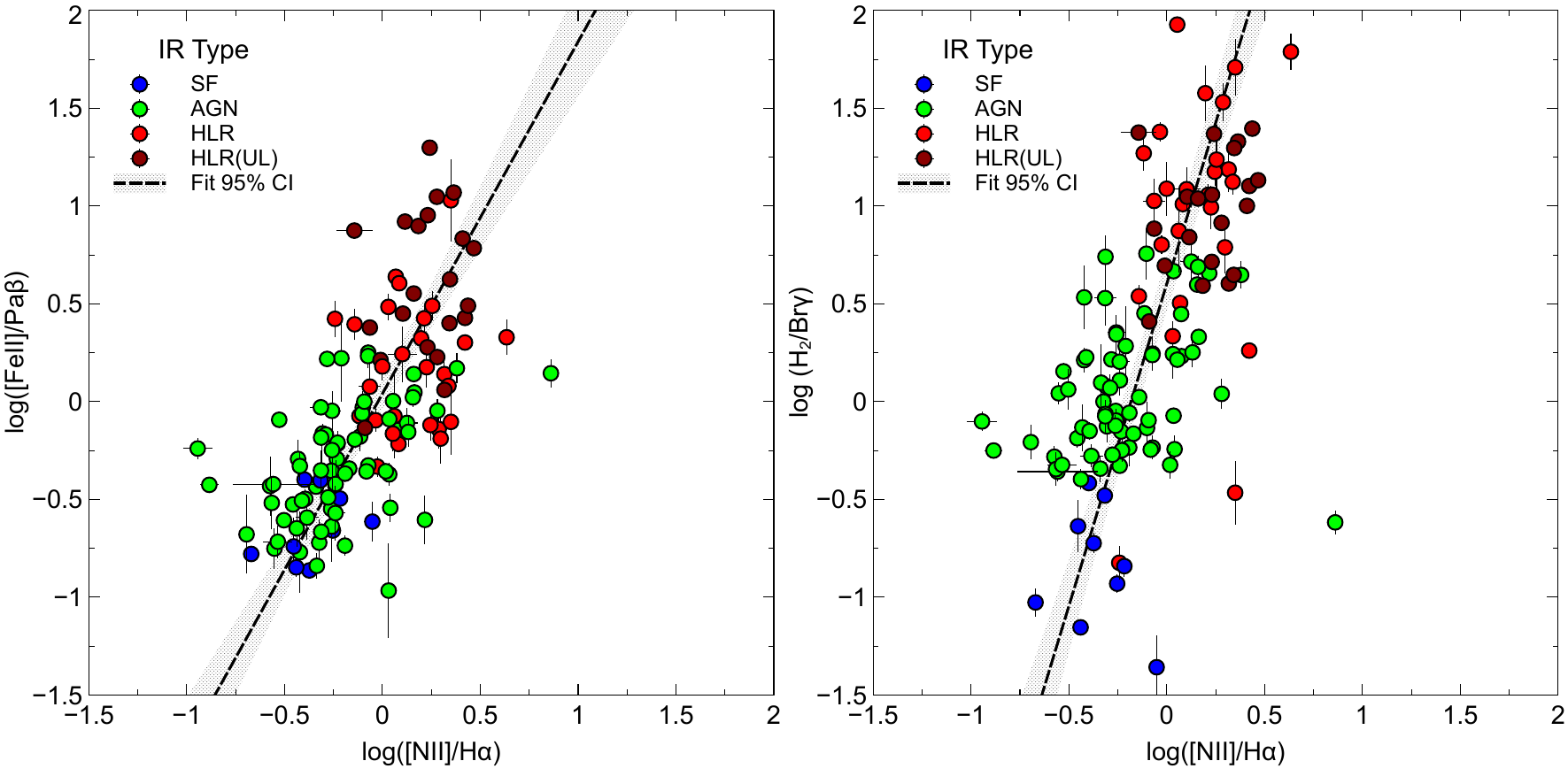}
    \caption{Plot of \NII/\Ha{} vs. \Fe/\pab, colour-coded by NIR type (left panel) and  \NII/\Ha{} vs. \Htwo/\brg{} (right panel), with linear regression fits with 95\% confidence intervals.}
    \label{fig:WPFPlots09}
\end{figure*}

We can also compare the diagnostic line ratios of \NII/\Ha{} vs. \Fe/\pab, as shown in Figure \ref{fig:WPFPlots09} (left panel). These are reasonably well correlated (except for some scatter with upper-limit measurements of \pab); the best fit is:
\begin{equation}
    \log(\rm{[Fe II]/Pa\beta}) = (1.8\pm0.22)\log(\rm{[N II]/H\alpha})- (0.04\pm0.05)
\end{equation}
The slope, being different from unity, indicates that the emission-line mechanism is not the same for the species \Fe{} and \NII. The source of the iron is from dust grains that are released by supernovae or other shocks \citep{Mouri2000}, which are subsequently photo-ionised, whereas the nitrogen is purely photo-ionised. 

For comparison, we also plot \NII/\Ha{} vs. \Htwo/\brg; these are again well correlated. This is expected from the data set, as \Fe/\pab{} is well correlated with \Htwo/\brg. Again, the mechanisms are very different; \Htwo{} emission is produced by a combination of X-ray heating, shocks and UV fluorescence. The fit is:
\begin{equation}
\begin{split}
    \log(\rm{H_{2}/Br\gamma}) = (3.29\pm0.44)\log(\rm{[N II]/H\alpha}) - (0.60\pm0.07)
\end{split}
\end{equation}

\begin{figure*}
    \centering
    \includegraphics[width=1\linewidth]{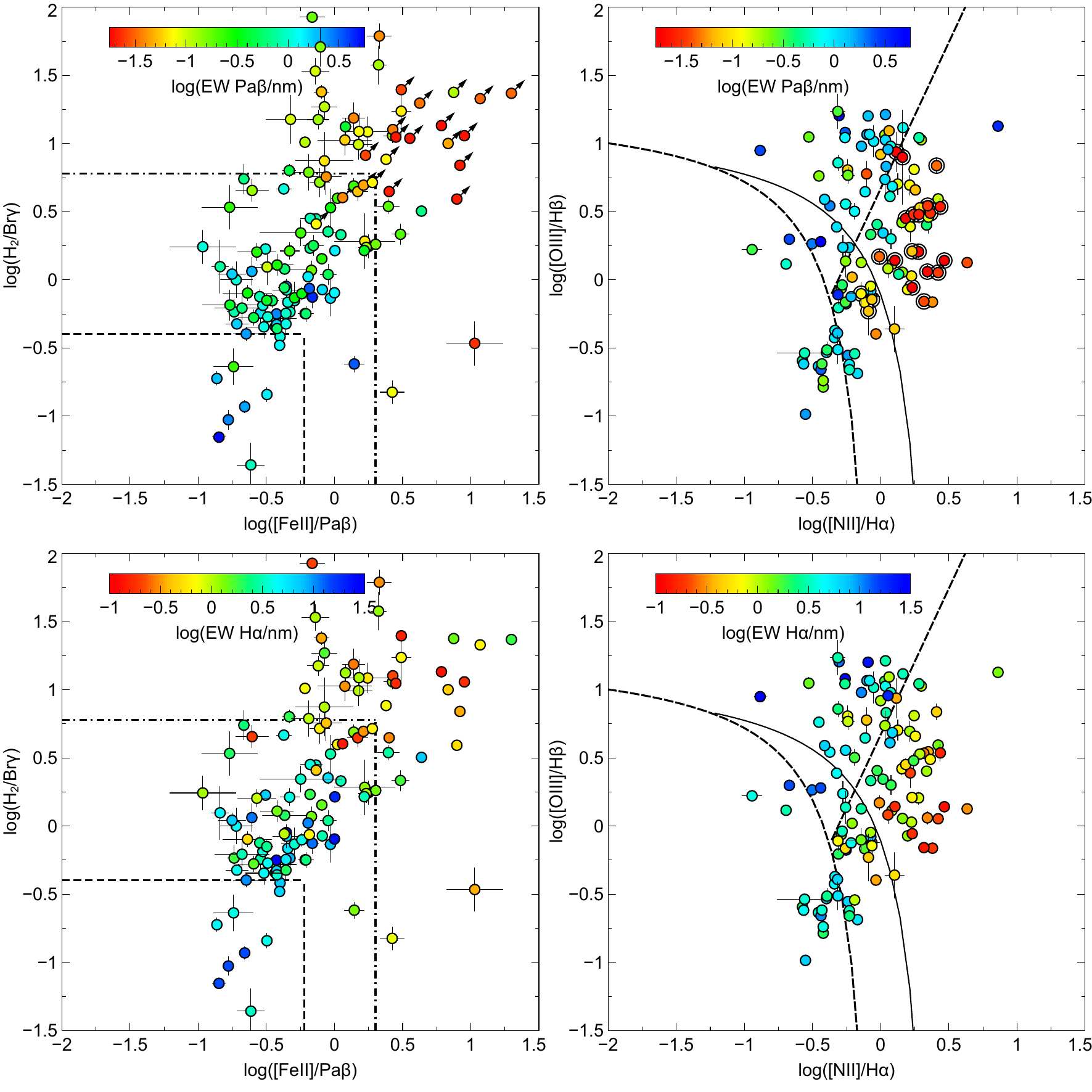}
    \caption{NIR and optical diagnostic diagrams, colour coded by \wpb{} (top row) and \wha{}(bottom row), showing strong trends with \wpb{} with IR and optical type; the \wha{} trends are less pronounced, but still clear. The NIR plot coded by \wpb{} has HLR(UL) types with arrows, and the same have circles on the points in the corresponding optical diagram.}
    \label{fig:WPFPlots03}
\end{figure*}

We also plot NIR and optical diagnostic diagrams (Figure \ref{fig:WPFPlots03}). These are colour coded by \wpb{} (top row) and \wha{}(bottom row). These show strong trends with \wpb{} with IR and optical type; the \wha{} trends are less pronounced, but still clear. The NIR plot coded by \wpb{} has HLR(UL) types with arrows, and the same objects have circles on the points in the corresponding optical diagram.
\subsection{The WPF Diagnostic}
\begin{figure}
    \centering
    \includegraphics[width=1\linewidth]{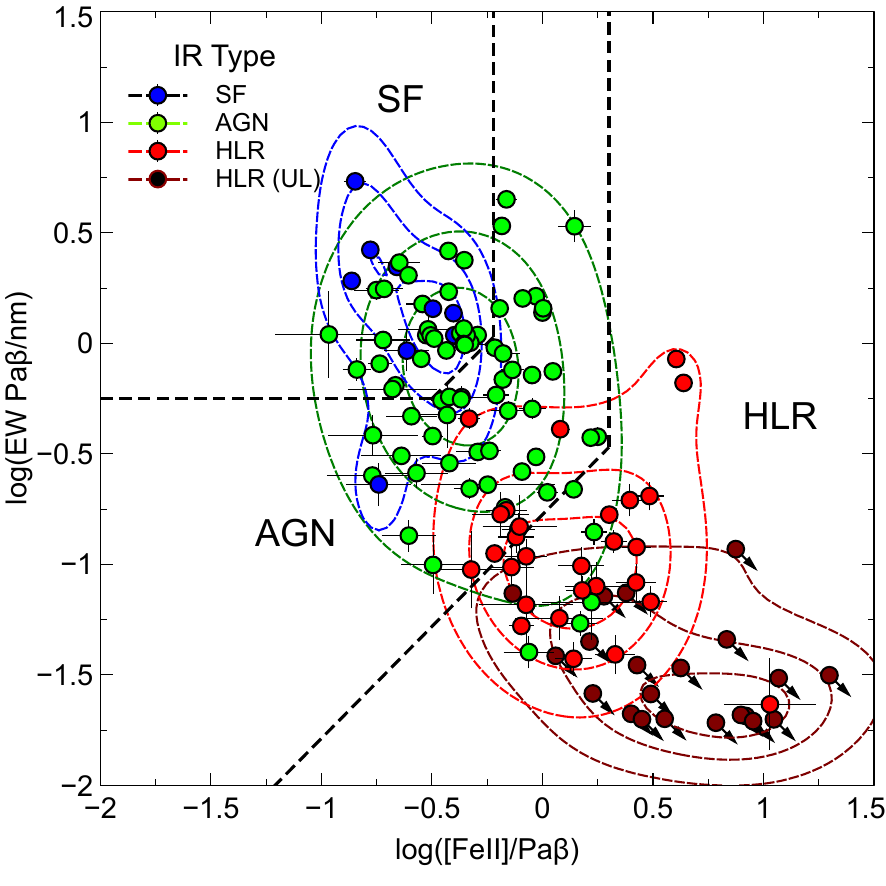}
    \caption{Proposed \wpb{} vs. \Fe/\pab{} diagram (WPF), showing SF, AGN and HLR regions, color-coded by the NIR classification. We note that the lack of pure SF galaxies to define cleanly the division between SF and AGN; this is due to the sample size.}
    \label{fig:WPFPlots05}
\end{figure}

Given the correlations between various indices as described above, we venture to introduce a new diagnostic, the \wpb{} vs. \Fe/\pab{} diagram (WPF - \textbf{W}idth of \textbf{P}aschen beta with \textbf{F}e). Figure \ref{fig:WPFPlots05} plots this diagram, colour-coded by NIR activity type; it can be seen that the activity types are reasonably disjoint. Tentative diagnostic regions have been drawn up; the boundaries are as follows:
\begin{enumerate}[label={\arabic*.},leftmargin=*,align=left]
\item Star forming galaxies lie above and to the left of the following lines.
\begin{flalign}
&\log(\Fe/Pa\beta) < -0.222 \\
&\nonumber\rm{and} \\
&\log(W_{Pa\beta}) > -0.25 \\
&\nonumber\rm{and} \\
&\log(W_{Pa\beta}) < \log(\Fe/Pa\beta) - 0.25
\end{flalign}
\item AGN are in the middle region of the diagram, defined by:
\begin{flalign}
&\log(\Fe/Pa\beta) >= -0.222 \\
&\nonumber\rm{or} \\
&\log(W_{Pa\beta}) <= -0.25 \\
&\nonumber\rm{or} \\
&\log(W_{Pa\beta}) >= \log(\Fe/Pa\beta) - 0.25
\end{flalign}
combined with:
\begin{flalign}
&\log(\Fe/Pa\beta) < 0.301\\
&\nonumber\rm{and} \\
&\log(\Fe/Pa\beta) > \log(W_{Pa\beta}) - 0.771
\end{flalign}
\item High line ratio objects are to the right and lower part of the diagram:
\begin{flalign}
&\log(\Fe/Pa\beta) => 0.301 \\
&\nonumber\rm{or} \\
&\log(\Fe/Pa\beta) <= \log(W_{Pa\beta}) - 0.771
\end{flalign}
\end{enumerate}
The $\log(\Fe/Pa\beta)$ divisions of -0.222 and 0.301 correspond with the standard NIR region for AGN of \Fe/\pab{} between 0.6 and 2 \citep{Riffel2013a}. It is noted that the SF region has few points (especially in the top left corner of the diagram) and it shows an overlap with the AGN region; this is due to having few genuine SF objects in the sample. We also show the classification matching between the standard NIR and WPF types (Table \ref{tbl:OIRD06}).

\begin{figure*}
    \centering
    \includegraphics[width=1\linewidth]{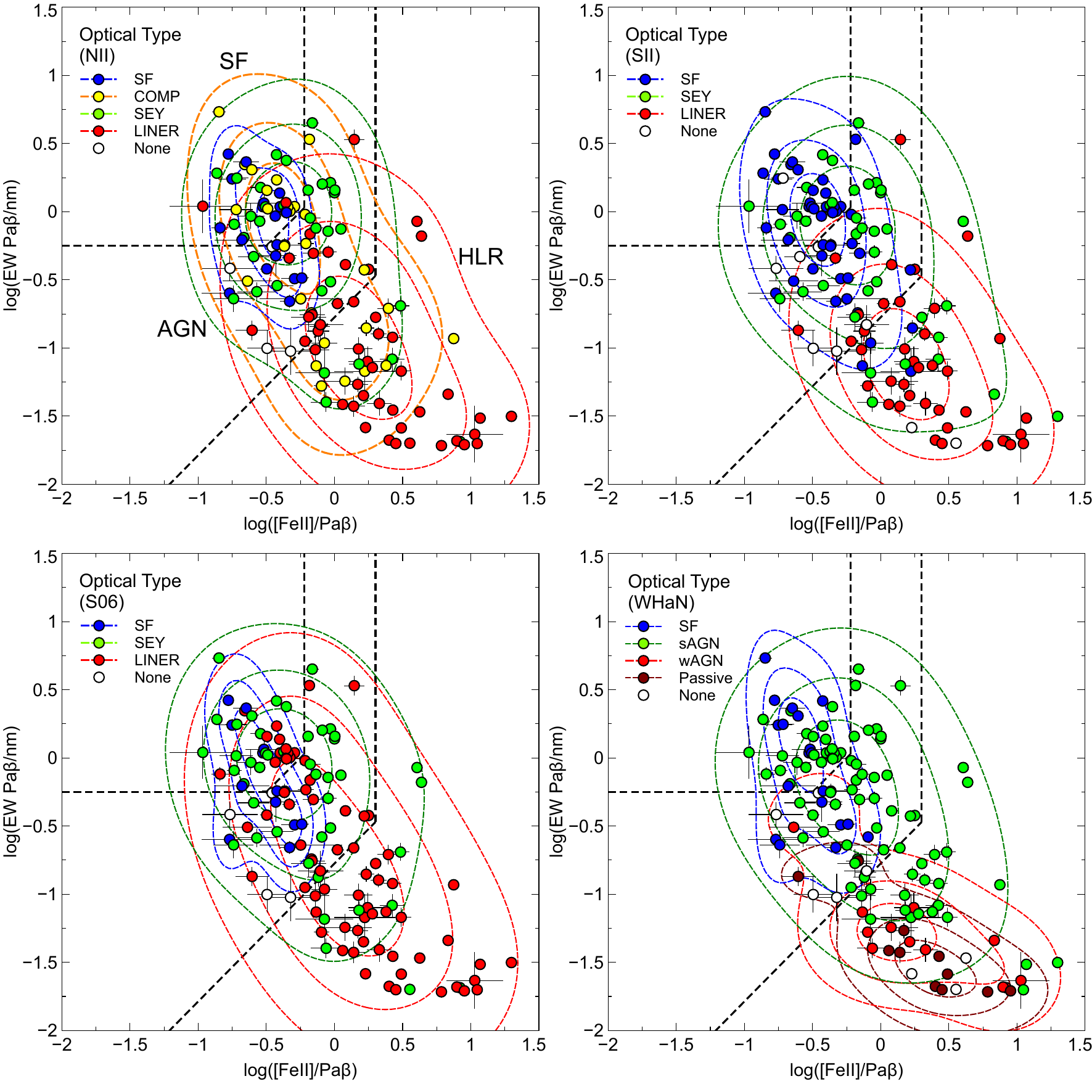}
    \caption{{The WPF diagnostic, compared to the optical types. Top left panel: \NII-\citetalias{Kauffmann2003}. Top right panel: \SII. Bottom left panel: \NII-\citetalias{Stasinska2006}. Bottom right panel: WHaN. These show large overlaps of SF into the AGN regime, with AGN and LINERs strongly overlapping for \NII-\citetalias{Stasinska2006}. There is minimal distinction between ``weak'' AGNs and passive galaxies in the WHaN diagram.}}
    \label{fig:WPFPlots06}
\end{figure*}

\begin{table}
\centering
\caption{Matching (in percentages of each standard classification) between the standard NIR (rows) and proposed WPF classification (columns).}
\label{tbl:OIRD06}
\begin{tabular}{lrrr}
\toprule
&\multicolumn{3}{c}{WPF Type}\\
NIR Type& SF   & AGN  & HLR   \\ \midrule
SF      & 85.7 & 14.3  & 0.0   \\
AGN     & 36.5 & 56.8 & 6.8   \\
HLR     & 0.0  & 26.8 & 71.4  \\
HLR(UL) & 0.0  & 0.0  & 100.0 \\ \bottomrule\end{tabular}
\end{table}
Figure \ref{fig:WPFPlots06} shows the WPF diagram, this time colour-coded by the optical indicators \NII-\citetalias{Kauffmann2003}, \NII-\citetalias{Stasinska2006}, \SII{} and WHaN. There is a large overlap of SF into the Seyfert region for the \NII-\citetalias{Kauffmann2003} and \SII{} diagrams; for the \NII-\citetalias{Stasinska2006} diagram, the LINER types also strongly overlap into the Seyfert region, again supporting the idea that the \NII-\citetalias{Stasinska2006} divisions between Seyfert and LINER are too strict. {The WHaN comparison shows that the passive and and retired objects cannot be well differentiated from the ``weak'' AGNs, with the ``strong'' AGNs showing a strong overlap in the WPF SF region.}

A limitation of this WPF classification is that it is based on only 132 galaxies, whereas the WHaN classification is built  on almost 125 000 galaxies from SDSS \citep{CidFernandes2011}. Until a large matched NIR survey can be conducted, our delineation of the WPF diagnostic is not definitive.

Table \ref{tbl:OIRD07} gives a sample of the results for each galaxy; Table \ref{tbl:OIRD07a} the NIR results and Table \ref{tbl:OIRD07b} the optical results. These show the object name, catalogue reference source (given in Tables \ref{tbl:OIRD01} and \ref{tbl:OIRD02}), line ratios and EWs, activity classifications and flags. The full table is available online; see the Data Availability section below.
\begin{subtables}
\label{tbl:OIRD07}
\begin{table*}
\caption{Sample of NIR flux ratios and activity classifications. (The full table is available online.)}
\label{tbl:OIRD07a}
\centering
\begin{tabular}{@{}lccccclll@{}}
\toprule
Object        & NIR Catalogue & log(\Fe/\pab)  & log(\Htwo/\brg) & log(EW \pab)   & UL Flag & NIR Type & WPF Type & IR Flags \\
(1)&(2)&&&&(3)&(4)&(5)&(6)\\
\midrule
1H1934-063    & 1          & -0.61$\pm$0.10 & -1.36$\pm$0.17  & -0.03$\pm$0.09 &         & SF      & SF       & 1        \\
ARK564        & 1          & -0.86$\pm$0.04 & -0.72$\pm$0.05  & 0.28$\pm$0.03  &         & SF      & SF       & 1        \\
ARP102B       & 1          & -0.33$\pm$0.05 & 0.80$\pm$0.05   & -0.34$\pm$0.05 & >>      & HLR     & AGN      & 3        \\
ESO428-14     & 1          & -0.09$\pm$0.03 & -0.07$\pm$0.02  & 0.20$\pm$0.01  &         & AGN     & AGN      & 6        \\
ESO507-25     & 3          & 0.08$\pm$0.20  & 1.03$\pm$0.11   & -1.24$\pm$0.10 &         & HLR     & HLR      &          \\
IC0537        & 3          & -0.07$\pm$0.09 & 1.27$\pm$0.09   & -0.96$\pm$0.08 & >>      & HLR     & HLR      & 3        \\
IC0630        & 3          & -0.85$\pm$0.05 & -1.15$\pm$0.03  & 0.73$\pm$0.02  &         & SF      & SF       &          \\
MCG+04-50-004 & 3          & -0.25$\pm$0.30 & 0.34$\pm$0.10   & -0.64$\pm$0.04 &         & AGN     & AGN      &          \\
MRK0124       & 1          & -0.54$\pm$0.07 & -0.24$\pm$0.07  & 0.18$\pm$0.03  &         & AGN     & SF       & 1        \\
MRK0291       & 1          & -0.72$\pm$0.12 & 0.00$\pm$0.13   & 0.02$\pm$0.04  &         & AGN     & SF       & 12       \\
MRK0334       & 1          & -0.29$\pm$0.02 & -0.15$\pm$0.04  & 0.04$\pm$0.01  &         & AGN     & SF       & 1        \\
MRK0493       & 1          & -0.74$\pm$0.05 & -0.23$\pm$0.10  & -0.09$\pm$0.03 &         & AGN     & SF       & 12       \\
MRK0509       & 1          & -0.74$\pm$0.15 & -0.64$\pm$0.13  & -0.64$\pm$0.10 &         & SF      & AGN      & 1        \\
NGC1052       & 2          & 0.25$\pm$0.05  & 0.25$\pm$0.09   & -0.42$\pm$0.04 &         & AGN     & AGN      &          \\
NGC1097       & 1          & 0.63           & 1.30            & -1.47          & >       & HLR(UL) & HLR      & 4        \\
NGC1144       & 1          & -0.19$\pm$0.13 & 0.79$\pm$0.13   & -0.77$\pm$0.10 &         & HLR     & AGN      &          \\
NGC1167       & 2          & -0.14$\pm$0.10 & 1.53$\pm$0.10   & -1.01$\pm$0.09 & >>      & HLR     & HLR      & 3        \\
NGC1174       & 4          & -0.40$\pm$0.04 & -0.48$\pm$0.03  & 0.14$\pm$0.02  &         & SF      & SF       &          \\
NGC1222       & 3          & -0.78$\pm$0.03 & -1.03$\pm$0.07  & 0.42$\pm$0.01  &         & SF      & SF       &          \\
NGC1275       & 1          & 0.61$\pm$0.04  & 2.69$\pm$0.34   & -0.07$\pm$0.04 &         & HLR     & HLR      & 1        \\
NGC1482       & 4          & -0.43$\pm$0.04 & -0.34$\pm$0.07  & -0.03$\pm$0.02 &         & AGN     & SF       &          \\
NGC1614       & 1          & -0.66$\pm$0.02 & -0.93$\pm$0.05  & 0.35$\pm$0.01  &         & SF      & SF       &          \\
NGC1635       & 3          & 0.21           & 0.69            & -1.35          & >       & HLR(UL) & HLR      & 4        \\
NGC1691       & 3          & -0.37$\pm$0.04 & -0.06$\pm$0.05  & -0.25$\pm$0.02 &         & AGN     & AGN      &     \\ \bottomrule
\end{tabular}
\begin{flushleft}
\small
\begin{tabular}{ll}
\textbf{Notes:}\\
1&Galaxy identifier.\\
2&Source of NIR spectral data, see Table \ref{tbl:OIRD01}.\\
3&Upper limit flag, ``>'' - \pab{} and \brg{} upper limits, ``>>'' - No \brg, computed frpm \pab{} value.\\
4&NIR activity type, as defined by \citet{Larkin1998, Rodriguez-Ardila2005, Riffel2013a,Riffel2021a}\\
5&WPF activity type, as defined in this work.\\
6&Flags - see Table \ref{tbl:OIRD08}\\
\end{tabular}
\end{flushleft}
\end{table*}
\begin{table*}
\caption{Sample of optical flux ratios and activity types. (The full table is available online.)}
\label{tbl:OIRD07b}
\centering
\resizebox{\textwidth}{!}{%
\begin{tabular}{lccccllllll}
\toprule
Object        & Opt. & log(\NII/\Ha)  & log(\SII/\Ha)  & log(\OIII/\Hb) & log(EW \Ha)    & Opt Type & Opt Type & Opt Type & Opt Type& Opt. \\ 
&Catalogue & & & & &\citetalias{Kauffmann2003}&\SII&\citetalias{Stasinska2006} &[WHaN]&Flags\\
&(1)&&&&&(2)&(3)&(4)&(5)&(6)\\
\midrule
1H1934-063    & 2  & -0.05$\pm$0.02 & -0.40$\pm$0.02 & 1.02$\pm$0.12  & 0.53$\pm$0.01  & SEY  & SEY & SEY & sAGN & 1 \\
ARK564        & 6  & -0.37$\pm$0.02 & -1.08$\pm$0.05 & 0.54$\pm$0.03  & 0.65$\pm$0.01  & SEY  & SF  & SEY & sAGN & 1 \\
ARP102B       & 2  & -0.03$\pm$0.01 & 0.12$\pm$0.01  & 0.41$\pm$0.02  & 0.29$\pm$0.01  & LIN  & LIN & LIN & sAGN &   \\
ESO428-14     & 8  & 0.04$\pm$0.02  & -0.18$\pm$0.02 & 1.21$\pm$0.01  & 0.48$\pm$0.01  & SEY  & SEY & SEY & sAGN &   \\
ESO507-25     & 1  & -0.06$\pm$0.06 & -0.04$\pm$0.07 & -0.12$\pm$0.09 & -0.51$\pm$0.05 & COMP & LIN & LIN & wAGN &   \\
IC0537        & 1  & -0.12$\pm$0.01 & -0.25$\pm$0.01 & -0.17$\pm$0.03 & 0.22$\pm$0.01  & COMP & SF  & LIN & sAGN &   \\
IC0630        & 11 & -0.44$\pm$0.02 & -0.78$\pm$0.02 & 0.28$\pm$0.01  & 1.16$\pm$0.02  & COMP & SF  & SEY & SF   &   \\
MCG+04-50-004 & 10 & -0.26$\pm$0.03 & -0.64$\pm$0.06 & 0.14$\pm$0.12  & 0.53$\pm$0.01  & COMP & SF  & LIN & sAGN &   \\
MRK0124       & 4  & 0.04$\pm$0.03  & -0.40$\pm$0.02 & 0.83$\pm$0.03  & 0.46$\pm$0.02  & SEY  & SEY & SEY & sAGN & 1 \\
MRK0291       & 4  & -0.32$\pm$0.02 & -0.55$\pm$0.01 & 0.39$\pm$0.01  & 0.65$\pm$0.01  & COMP & SF  & SEY & sAGN & 1 \\
MRK0334       & 8  & -0.23$\pm$0.01 & -0.67$\pm$0.01 & 0.24$\pm$0.00  & \dots          & COMP & SF  & LIN & sAGN &   \\
MRK0493       & 4  & -0.19$\pm$0.02 & -0.44$\pm$0.01 & 0.50$\pm$0.06  & 0.25$\pm$0.01  & SEY  & SEY & SEY & sAGN & 1 \\
MRK0509       & 2  & -0.45$\pm$0.02 & -1.02$\pm$0.02 & 0.76$\pm$0.02  & 0.60$\pm$0.01  & SEY  & SEY & SEY & SF   & 1 \\
NGC1052       & 2  & -0.07$\pm$0.02 & -0.03$\pm$0.02 & 0.33$\pm$0.02  & 0.44$\pm$0.01  & LIN  & LIN & LIN & sAGN &   \\
NGC1097       & 15 & 0.34           & -0.06$\pm$0.01 & 0.54$\pm$0.00  & \dots          & LIN  & LIN & LIN &      & 8 \\
NGC1144       & 7  & 0.30$\pm$0.01  & 0.05$\pm$0.02  & 1.03$\pm$0.04  & -0.01$\pm$0.00 & LIN  & SEY & SEY & sAGN &   \\
NGC1167       & 3  & 0.29$\pm$0.01  & 0.18$\pm$0.02  & 0.53$\pm$0.02  & 0.01$\pm$0.01  & LIN  & LIN & LIN & sAGN &   \\
NGC1174       & 3  & -0.32$\pm$0.01 & -0.56$\pm$0.01 & -0.39$\pm$0.02 & 0.82$\pm$0.01  & SF   & SF  & LIN & sAGN &   \\
NGC1222       & 1  & -0.67$\pm$0.02 & -0.68$\pm$0.01 & 0.30$\pm$0.01  & 1.16$\pm$0.00  & SF   & SF  & SF  & SF   &   \\
NGC1275       & 2  & 0.09$\pm$0.02  & -0.16$\pm$0.02 & 0.69$\pm$0.01  & 0.79$\pm$0.01  & LIN  & SEY & SEY & sAGN &   \\
NGC1482       & 1  & -0.34$\pm$0.01 & -0.49$\pm$0.01 & -0.42$\pm$0.04 & 0.66$\pm$0.00  & SF   & SF  & LIN & sAGN &   \\
NGC1614       & 1  & -0.25$\pm$0.01 & -0.66$\pm$0.02 & -0.18$\pm$0.01 & 1.17$\pm$0.00  & COMP & SF  & LIN & sAGN &   \\
NGC1635       & 4  & -0.01$\pm$0.03 & -0.17$\pm$0.05 & 0.17$\pm$0.06  & -0.51$\pm$0.02 & LIN  & LIN & LIN & wAGN & 2 \\
NGC1691       & 13 & -0.19$\pm$0.01 & -0.51$\pm$0.01 & -0.54$\pm$0.03 & -0.01$\pm$0.00 & COMP & SF  & LIN & sAGN & 9\\        
\bottomrule\\
\end{tabular}
}
\begin{flushleft}
\small
\begin{tabular}{ll}
\textbf{Notes:}\\
1&Source of optical spectral data, see Table \ref{tbl:OIRD02}.\\
2,3,4,5&Optical activity classification, see text.\\
6&Flags - see Table \ref{tbl:OIRD08}\\
\end{tabular}
\end{flushleft}
\end{table*}
\end{subtables}
\begin{table*}
\caption{NIR and optical flag values for Table \ref{tbl:OIRD07}}
\label{tbl:OIRD08}
\begin{flushleft}
\begin{tabular}{ll}
\toprule
\multicolumn{2}{l}{\textbf{NIR Flags for Table \ref{tbl:OIRD07a}}}\\
\midrule
1 & Broad hydrogen lines modelled with two components. \\
2 & Broad hydrogen lines modelled with Lorentzian curve. \\
3 & No \brg{} measurement, ratio from \pab{} used.\\
4 & No hydrogen lines (\pab{} or \brg), upper limits used.\\
5 & Hydrogen line fits uncertain, due to noise or multiple components.\\
6 & Outflows, modelled with 2 Gaussian curves.\\
\midrule
\multicolumn{2}{l}{\textbf{Optical Flags for Table \ref{tbl:OIRD07b}}}\\ \midrule
1 & Broad hydrogen lines, fitted with two components. \\
2& No \Hb{} measurement, ratio from \Ha{} used.\\
3& No \OIII{} measurement.\\
4&\SII{} outside of spectral range or no \SII{} flux measurable.\\
5&No blue spectrum available.\\
6&\OIII{}$\lambda$5007 \AA{} out of spectral range, ratio from \OIII{}$\lambda$4959 \AA{} used.\\
7&Spectrum too noisy.\\
8&Flux values acquired from tabular data.\\
9&Digitised plots from the reference source paper.\\
\bottomrule
\end{tabular}
\end{flushleft}
\end{table*}
\section{Conclusions}
\label{sec:OIRD_5}
We have studied a catalogue of 132 emission-line galaxies \citep[combined from the catalogues of][]{Riffel2006, Mason2015, Durre2023,Martins2013} with matched NIR and optical spectra, to examine the relationship between the respective activity classifications, which are deduced by ratios of the fluxes of emission lines. We compare the standard NIR classification \citep{Larkin1998, Rodriguez-Ardila2005,Riffel2013a,Riffel2021a} with three optical classifications \citep{Kewley2001a, Kauffmann2003, Stasinska2006} plus the WHaN classification of \cite{CidFernandes2010a,CidFernandes2011}. The conclusions are as follows:
\begin{enumerate}[label={\arabic*.},leftmargin=*,align=left]
    \item  While there is a broad match between the two regimes (i.e. NIR and optical), there are mismatches and overlaps caused by either (i) aperture effects from the disparate observational setups and/or (ii) NIR radiation penetrating obscuring dust and ``seeing deeper'' into the nuclear region, exposing AGN activity. {In general, it would be unsafe to predict the IR type from the optical type(s) and vice versa.}
    \item The \NII-\citetalias{Stasinska2006} classification produces the largest mismatch to the NIR types; this is considered to be because the region boundaries of \NII-\citetalias{Stasinska2006} are too strict.
    \item The COMP classification of  the \NII-\citetalias{Kauffmann2003} diagram is not visible in the NIR types and can be subsumed into either SF or AGN activity types.
    \item We examined the relationship between the EWs of \Ha{} and \pab, as well as the ratios \NII/\Ha{} vs. \Fe/\pab, and find reasonable correlations.
    \item We can thus propose a new diagnostic (WPF) in the NIR,  analogous to the WHaN classification, using the \Fe/\pab{} flux ratio and the EW of the \pab{} line. We show, within the limitations of the catalogue size, that the SF, AGN and HLR classifications (from the standard NIR diagram) can be reasonably replicated in this scheme. This diagnostic has the advantage that only one wavelength range (\textit{J}-band spectra for low \textit{z} objects) needs to be observed, thus being economical with telescope time.
\end{enumerate}
We acknowledge the limitations of this survey, with a bias against ``true'' SF types in the NIR, and too low numbers to draw definitive delineations of the proposed WPF classification. {Any future survey has to carefully consider aperture effects, ideally observing with one instrument for the complete optical and NIR spectral regions simultaneously (e.g. X-Shooter on VLT).}

{As a further comment, it is very desirable that spectral surveys publish and archive the reduced spectra, which are an invaluable resource for future work. Several sources in this work conducted large optical surveys and published results without the actual spectra, with the raw observations no longer accessible. In these days of large data archives and organisations like CDS, this is no longer acceptable.}
\section*{Data Availability}
Table \ref{tbl:OIRD07} is available in full at CDS via anonymous ftp to \url{cdsarc.u-strasbg.fr} (\url{130.79.128.5}) or via \url{https://cdsarc.unistra.fr/viz-bin/cat/J/MNRAS/???/???}.

\bibliographystyle{mnras}
\bibliography{library} 

\bsp	
\label{lastpage}
\end{document}